\documentclass[10pt,conference]{IEEEtran}
\IEEEoverridecommandlockouts
\usepackage[letterpaper, top=0.75in, bottom=1.037in, left=0.624in, right=0.624in]{geometry}
\usepackage{gensymb}
\usepackage{cite}
\usepackage{amsmath,amssymb,amsfonts}
\usepackage{algorithmic}
\usepackage{graphicx}
\usepackage{textcomp}
\usepackage{xcolor}
\usepackage{threeparttable}
\usepackage{orcidlink}
\usepackage[nameinlink,capitalize]{cleveref}
\usepackage{balance}
\crefname{equation}{}{}
\crefname{section}{Sec.}{Secs.}
\crefname{figure}{Fig.}{Figs.}
\usepackage{tcolorbox}
\usepackage{booktabs,tabularx,siunitx,array}
\definecolor{skyblue}{RGB}{135, 206, 235}
\definecolor{lightyellow}{RGB}{255, 255, 128}

\usepackage[lang=en]{jabbrv}

\def\BibTeX{{\rm B\kern-.05em{\sc i\kern-.025em b}\kern-.08em
    T\kern-.1667em\lower.7ex\hbox{E}\kern-.125emX}}

\makeatletter
\newcommand{\linebreakand}{%
  \end{@IEEEauthorhalign}
  \hfill\mbox{}\par
  \mbox{}\hfill\begin{@IEEEauthorhalign}
}
\makeatother
\begin{document}
\bstctlcite{IEEEexample:BSTcontrol}

\title{Keys in the Weights: Transformer Authentication Using Model-Bound Latent Representations}

\author{\IEEEauthorblockN{Ayşe~S.~Okatan, Mustafa~İlhan~Akbaş\orcidlink{0000-0002-5450-3522}, Laxima~Niure~Kandel
and Berker~Peköz\orcidlink{0000-0002-7572-3663}
\thanks{Code and data available at: \url{https://github.com/maverai/Keys-in-the-Weights}}
}
\IEEEauthorblockA{
Dept. of Electrical Engineering and Computer Science, Embry-Riddle Aeronautical University,
Daytona Beach, FL, USA\\
e-mail: \textit{\href{mailto:okatana@my.erau.edu}{okatana@my.erau.edu}}, \{\textit{\href{mailto:akbasm@erau.edu}{akbasm}},\textit{\href{mailto:Laxima.NiureKandel@erau.edu}{ Laxima.NiureKandel}},\textit{\href{mailto:Berker.Pekoz@erau.edu}{Berker.Pekoz}}\}\textit{@erau.edu}
}}
\maketitle

\begin{abstract}
We introduce \emph{Model-Bound Latent Exchange (MoBLE)}, a decoder-binding property in Transformer autoencoders formalized as \emph{Zero-Shot Decoder Non-Transferability (ZSDN)}. In identity tasks using iso-architectural models trained on identical data but differing in seeds, self-decoding achieves $> 91\%$ exact match and $> 98\%$ token accuracy, while zero-shot cross-decoding collapses to chance ($\approx 1/$vocabulary size) without exact matches. This separation arises without injected secrets or adversarial training, and is corroborated by weight-space distances and attention-divergence diagnostics. We interpret ZSDN as \emph{model binding}--a latent-based authentication and access-control mechanism--even when the architecture and training recipe are public%
: encoder's hidden state representation deterministically reveals the plaintext, yet only the correctly keyed decoder reproduces it in zero-shot%
. We formally define ZSDN, a decoder-binding advantage metric, and outline deployment considerations for secure artificial intelligence (AI) pipelines. Finally, we discuss learnability risks (e.g., adapter alignment) and outline mitigations. MoBLE offers a lightweight, accelerator-friendly approach to \emph{secure AI deployment} in safety-critical domains, including aviation and cyber-physical systems.
\end{abstract}


\section{Introduction}
AI systems are increasingly deployed in safety-critical environments where model integrity and controlled interoperability are vital. While cryptographic protocols protect data confidentiality, they do not address a complementary question: 
\emph{Are latent representations produced by one model without secrets or cryptographic machinery decodable only by its paired decoder? Can decoder output alone authenticate the encoding transformer?}

 This question is central to secure model-to-model communication, provenance, and AI safety.
In principle, any two independently trained neural networks, such as recurrent (long-short term memory (LSTM)/gated recurrent unit (GRU)) (RNN), convolutional (CNN) or multilayer perceptron (MLP) that do a lossless compression can end up with different internal encodings (different "keys") identifying that network \cite{kanter2002secure,klimov2002analysis,abadi2017learning,Akinci2024}. 

However, RNNs share parameters across time, CNNs share filters across positions; these weight-sharing constraints reduce the independent degrees of freedom, making collisions or partial alignments more likely (and thus less secure in a cryptographic sense). For the RNN/CNN/MLP architectures to exhibit this effect requires explicit architectural enforcement (e.g., an enforced compression and removal of trivial identity solutions) to observe such behaviors robustly. An additional effort is needed to incorporate explicit secrets or adversarial elements in training to obtain distinguishable identity in these architectures.

Transformer architectures \cite{vaswani2017attention}, on the other hand, amplify this effect intrinsically because of their high expressiveness and multiple equivalent solutions stemming from attention layers. Attention mechanism of transformers creates multiple plausible encoding functions for the same task, effectively giving each transformer model an individualized encoder. Transformers have been studied chiefly for privacy-preserving inference \cite{zheng2023primer,moon2024thor,zhang2024nexus,hao2023iron} and model locking/watermarking in vision \cite{kiya2023blockwise,kiya2023transformation}. 

A long line of work has asked whether independently trained networks learn equivalent internal representations up to affine or orthogonal transforms \cite{Li2015ConvergentLearning} and how to test this via singular vector canonical component analysis\cite{raghu2017svccasingularvectorcanonical} or centered kernel alignment\cite{kornblith2019similarity} or by model stitching \cite{hernandez2023modelstitchinglookingfunctional}. Stitching succeeds only after learning a small connector between models\cite{pan2023stitchable}, suggesting compatibility is not free\cite{bansal2021revisitingmodelstitchingcompare}.
We study the complementary regime: independently trained transformer autoencoders--identical in architecture, tokenizer, hyperparameters and training data but initialized with different random seeds--learn \emph{non-transferable latent spaces}. Specifically, an encoder's final memory $H^L$ is reliably decodable only by its own decoder; zero-shot cross-decoding by another model collapses to chance-level token accuracy ($\approx 1/|V|$) and 0\% exact matches, despite architectural symmetry. This phenomenon, which we term \emph{Zero-Shot Decoder Non-Transferability (ZSDN)}, emerges naturally from seed-induced basis misalignment in attention projections and feed-forward layers. Interpreting the learned weights as private key material places our construction within the Kerckhoffs–Shannon tradition of public algorithms with secret keys, while remaining distinct from model watermarking\cite{Uchida_2017,adi2018turning}/locking mechanisms \cite{Rouhani2019DeepSigns}. We further include two controls: an exact post-training clone of M1 (bit-identical weights) and a fresh re-training with the same seed to probe determinism effects \cite{pytorch_determinism}. Our contributions can be summarized as follows:


\begin{itemize}
    \item We formalize ZSDN and \emph{decoder-binding advantage} metric quantifying the gap between self- and cross-decoding.
    \item We support the basis-misalignment hypothesis using weight-space metrics and attention-divergence diagnostics.
    \item We reposition this parameter identity as \emph{model binding} for access control and authentication, discussing learnability risks (e.g., adapter attacks) and outlining mitigations.
    \item We propose a deployment checklist (e.g., quantization, integrity tags, operational key rotation), framing MoBLE as a lightweight security layer for AI pipelines in aviation.
\end{itemize}
\textbf{Scope.} Our findings are based on a character-level identity task with small Transformer autoencoders; generalization to larger models and non-identity tasks remains future work. Nevertheless, the sharp zero-shot failure of cross-decoding under identical training recipes highlights a structural property of Transformer parameterization for secure AI deployments.

The remainder of this article is organized as follows: \Cref{sec:background} reviews related work. \Cref{sec:method} details our data, architecture, training, and evaluation protocol. \Cref{sec:results} presents quantitative results and attention diagnostics.
\Cref{sec:conc} concludes with implications for secure model-to-model communication.

\section{Related Work}
\label{sec:background}

\subsection{Neural cryptography}
Early "neural key exchange" schemes showed that tree-parity machines can synchronize weights over a public channel to share a secret\cite{kanter2002secure}, though practical cryptanalytic attacks followed\cite{klimov2002analysis}. A modern line casts encryption/decryption as an adversarial learning game (Alice/Bob vs.\ Eve), demonstrating learned private-key behavior with vanilla primitives\cite{abadi2017learning}. Our setting uses \emph{no} adversarial loss or protocol, specificity emerges purely from independent public training trajectories.

\subsection{Transformers in Security and Privacy}
Transformers~\cite{vaswani2017attention} have been studied under encryption and data/model perturbations. Key-tied image/model transformations\cite{kiya2023transformation} (e.g., block-wise encryption\cite{kiya2023blockwise}) that effectively "lock" utility to secrets are explored for vision. Output watermarking embeds verifiable statistical signatures for language provenance~\cite{kirchenbauer2023watermark,kirchenbauer2024reliability}. These works target ownership/robustness for a \emph{single} model. By contrast, we study \emph{inter-model} specificity: encodings from one independently trained Transformer systematically fail to decode on another, despite identical architecture and training data.

\subsection{Privacy-Preserving Transformer Inference}
A complementary body of work runs Transformers over protected data using cryptography or trusted execution. Examples include HE-based and hybrid systems for BERT/LLMs (e.g., Primer~\cite{zheng2023primer}, THOR~\cite{moon2024thor}, NEXUS~\cite{zhang2024nexus}) and systems work on private Transformer inference~\cite{hao2023iron,rho2025efla}; recent surveys contextualize these directions~\cite{li2025ptiSurvey}. These approaches protect \emph{data confidentiality} for a single model provider. Our goal is orthogonal: we show that iso-architectural models trained from different seeds do \emph{not} interoperate at the level of latent memory $H^L$, yielding a natural key-mismatch failure mode.

\subsection{Model-Level Security and Key-Like Behavior}
Closest to our setting are "model locking" methods for ViTs, where secret transforms on data and/or parameters gate accuracy~\cite{kiya2023transformation,kiya2023blockwise}. Our empirical contribution is that even \emph{without} injected secrets, seed-driven optimization induces latent spaces that act like private keys—encodings decode only with the originating weights. This is distinct from output watermarking~\cite{kirchenbauer2023watermark,kirchenbauer2024reliability}; we leverage attention’s emergent non-transferability as a security primitive rather than defend against it.

\subsection{Geometry of Solutions and Weight-Space Connectivity}
Despite different initializations, many optima are connected by low-loss curves in parameter space \cite{garipov2018losssurfacesmodeconnectivity}, sometimes even linearly once permutation symmetries are accounted for \cite{entezari2022rolepermutationinvariancelinear}. Weight averaging within a basin ("model soups") can improve performance without extra inference cost \cite{wortsman2022model}. These results address \emph{path connectivity in weight space}; our phenomenon concerns \emph{interface compatibility at a fixed latent}, where minor weight changes can still render zero-shot decoding inoperable.

\subsection{Neural Encoders for Communication}
End-to-end learned communication uses autoencoders to co-design modulation and decoding under channel models~\cite{oshea2017physical}. Conceptually our pipeline is also encoder--decoder, but the "channel" is another model’s decoder. The \emph{failure} of cross-decoding between independently trained, iso-architectural Transformers is precisely the cryptographic property we exploit.

Prior work secures \emph{data} for a given model or ties a model to an \emph{explicit} secret. We instead show that independently trained Transformers already induce \emph{keyed} latent spaces: the encoder’s representation functions as a private key that only the identically parameterized decoder can invert.

\section{Methodology}\label{sec:method}

\subsection{Setup and Threat Model}
Let $\{f_j\},j\in\mathbb{N}$ denote Transformer encoder--decoder models with \emph{identical} architecture, tokenizer, and training data.
In our setting, the encoder of Model~1 (M1) produces a latent representation of a plaintext message $M$, while the decoders of other models $f_j$ ($j\neq 1$) act as adversarial receivers attempting to decode this representation without access to the originating initialization.
Models differ \emph{only} by random seed, yielding divergent learned parameters $\Theta^{Q,K,V}_j$ after training.

\begin{figure*}[t]
  \centering
  \includegraphics[width=0.95\linewidth,
                   trim=12mm 72mm 12mm 72mm,clip]{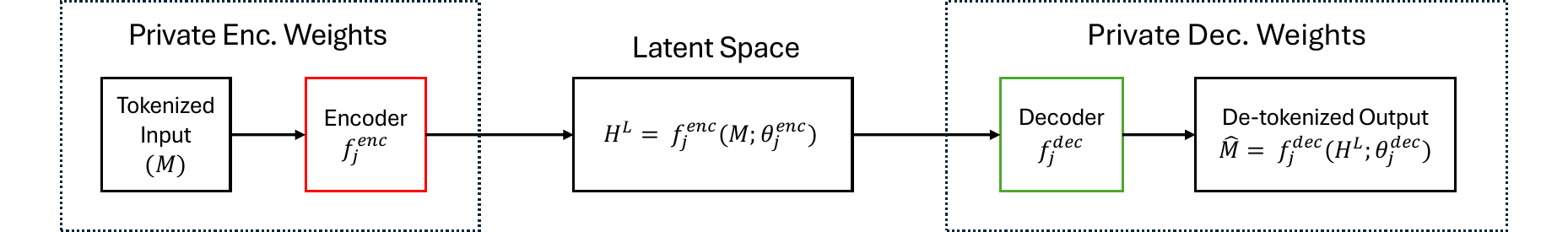}
  \caption{Authentication perspective. \textit{Note—} We adopt the neural–cryptography
  view that the model’s learned parameters act as an implicit private key:
  the encoder maps tokenized plaintext \(M\) to a latent \(H^{L}\), and only a
  decoder with the matching parameters reliably reconstructs \(\hat M\)
  \cite{kanter2002secure,klimov2002analysis,abadi2017learning}.}
  \label{fig:key}
\end{figure*}

\subsection{Vocabulary and Data Generation}
We use a character-level vocabulary with $|V|{=}86$ tokens: three specials $\{\langle\mathrm{pad}\rangle,\langle\mathrm{bos}\rangle,\langle\mathrm{eos}\rangle\}$, $26$ lowercase, $26$ uppercase, $10$ digits, and $21$ symbols
$\{\text{space},\; .,\,,:\,;\,!\,?,-,\_,/,+,*,=,(,),[,],\{,\},@,\#\}$. Vocabulary size also determines the integer-to-string cardinality. 
We synthesize an \emph{identity} corpus (input equals target): training uses $6{,}000$ sequences and testing $800$ sequences, each with length $L\sim\mathrm{Unif}[8,30]$ from the token pool excluding specials.
Sequences are encoded as $[\langle\mathrm{bos}\rangle, s_1,\ldots,s_\ell,\langle\mathrm{eos}\rangle]$ and truncated to $T_{\max}=50$.
Batches are padded to the \emph{batch-local} maximum length using $\langle\mathrm{pad}\rangle$ special token which is masked in attention and ignored in the loss.

\subsection{Architecture and Notation}
All models share: $d_{\text{model}}=256$, $L=4$ encoder--decoder layers, $h=4$ heads, $d_{\text{ff}}=1024$, dropout $=0.1$, and $T_{\max}=50$.
Token embeddings (separate source/target) are added to fixed sinusoidal positional encodings~\cite{vaswani2017attention}.
For layer $l$ and head $i$ (with $d_k=d_{\text{model}}/h$), the encoder’s multi-head self-attention~\cite{vaswani2017attention} uses
\begin{equation*}
\mathbf{Q}^{(l,i)}_{j} = \mathbf{X}^{(l)} \mathbf{W}^{Q,(l,i)}_{j}, 
\quad
\mathbf{K}^{(l,i)}_{j} = \mathbf{X}^{(l)} \mathbf{W}^{K,(l,i)}_{j}, 
\end{equation*}
\begin{equation*}
\mathbf{A}^{(l,i)}_{j} = \mathrm{softmax}\!\left(\frac{\mathbf{Q}^{(l,i)}_{j}\, \mathbf{K}^{(l,i)\top}_{j}}{\sqrt{d_k}}\right)
\in \mathbb{R}^{T\times T},
\end{equation*}
\begin{equation*}
\mathbf{V}^{(l,i)}_{j} = \mathbf{X}^{(l)} \mathbf{W}^{V,(l,i)}_{j}, \quad
\mathbf{Z}^{(l,i)}_{j} = \mathbf{A}^{(l,i)}_{j}\, \mathbf{V}^{(l,i)}_{j},
\end{equation*}
\begin{equation*}
\mathbf{Z}^{(l)}_{j} = \mathrm{Concat}\!\big(\mathbf{Z}^{(l,1)}_{j},\dots,\mathbf{Z}^{(l,h)}_{j}\big)\,
\mathbf{W}^{O,(l)}_{j}. \label{eq:mhaconcat}
\end{equation*}
Sublayers are residually connected and LayerNormalized\cite{ba2016layernormalization}:
\begin{align}
\tilde X^{(l)}_{j} &= \mathrm{LN}\!\big(X^{(l)}_{j} + \mathrm{Drop}(\mathrm{MHA}(X^{(l)}_{j}))\big),\label{eq:ln1}\\
X^{(l+1)}_{j} &= \mathrm{LN}\!\big(\tilde X^{(l)}_{j} + \mathrm{Drop}(\mathrm{FFN}(\tilde X^{(l)}_{j}))\big),\label{eq:ln2}
\end{align}
where $\mathrm{FFN}(x)=W_2\,\phi(W_1 x)$ with $\phi=\mathrm{GELU}$~\cite{hendrycks2016gelu}.
The decoder mirrors this structure and additionally uses (i) a causal mask $S_{tt'}=\mathbb{1}[t'\le t]$ in its self-attention and (ii) a cross-attention block over the encoder memory~\cite{vaswani2017attention}.
Finally, a linear generator maps decoder states to $|V|$ logits.

\subsection{Training Objective and Optimization}
We train with teacher forcing~\cite{williams1989learning} to minimize the next-token negative log-likelihood (NLL)
\begin{equation}
\mathcal{L}=-\sum_{t=1}^{x_{t}\neq \langle\mathrm{pad}\rangle}\log p_\theta(x_{t+1}\mid x_{\le t}),\label{eq:nll}
\end{equation}
 on decoder outputs aligned to the shifted target. Pairwise weight deviation is
\begin{equation}
D_{\ell_2}(a,b)=\Big(\sum_{p\in\Theta}\|p_a-p_b\|_2^2\Big)^{1/2}.\label{eq:l2}
\end{equation}
Optimization uses AdamW~\cite{loshchilov2019adamw} (learning rate $3{\times}10^{-4}$, weight decay $0$), batch size $128$, $8$ epochs, and global gradient clipping at $1.0$~\cite{pascanu2013difficulty}.
Dropout is applied at rate $0.1$~\cite{srivastava2014dropout}.
Both $\texttt{cuda}$ and $\texttt{cpu}$ devices are tested, demonstrating that results transfer across processing hardware architectures.
We instantiate five models: M1 (seed 111), M2 (222), M3 (333), M1\_CLONE (deep copy of M1), M1\_SAMESEED (training seed 111 on different device).

\subsection{Decoding and Cross-Decoding Protocol}
For self-decoding, we compute the encoder memory $H^L$ and greedily decode from $\langle\mathrm{bos}\rangle$, appending $\arg\max$ tokens until the first $\langle\mathrm{eos}\rangle$ in the batch or $T_{\max}$.
For cross-decoding, we \emph{fix} the memory $H^L$ and source key-padding mask produced by encoder $a$, and run decoder $b$ (its own parameters) using the same greedy procedure.
This mirrors the implementation in \texttt{decode\_with\_external\_memory} and keeps the latent fixed while swapping decoders.

\subsection{Attention Diagnostics}
To probe model specificity, we extract \emph{layer-0} attention maps only (as implemented) and average over heads to obtain $\bar{A}\in\mathbb{R}^{T\times T}$, row-normalized within $\varepsilon=10^{-9}$.
Given two models on the same inputs, we compute the mean (row-wise) Kullback–Leibler divergence~\cite{kullback1951information}
\begin{equation}
\mathrm{KL}(a\Vert b)
=\mathbb{E}_{t}\!\Big[\sum_{t'} \bar{A}^{(a)}_{t,t'} \log \tfrac{\bar{A}^{(a)}_{t,t'}}{\bar{A}^{(b)}_{t,t'}}\Big],\label{eq:kl}
\end{equation}
and the cosine similarity between flattened maps
\begin{equation}
\mathrm{Cos}(a,b)
=\frac{\langle \mathrm{vec}(\bar{A}^{(a)}),\mathrm{vec}(\bar{A}^{(b)})\rangle}
{\|\mathrm{vec}(\bar{A}^{(a)})\|_2\,\|\mathrm{vec}(\bar{A}^{(b)})\|_2}.\label{eq:cos}
\end{equation}
Decoder diagnostics follow the same pattern, capturing (i) decoder self-attention at the final step and (ii) decoder cross-attention to the fixed encoder memory under true cross-decoding.

\subsection{Evaluation Batches and Metrics}
To ensure pairwise comparability, we pre-fetch the first $6$ mini-batches from the test loader and reuse them across all ordered (encoder$\to$decoder) pairs.
For hypothesis $\hat{s}$ and reference $s$ we report:
\begin{itemize}
\item \textbf{Exact match (\%)}: $\mathbb{1}[\hat{s}=s]$, averaged over samples.
\item \textbf{Token accuracy (\%)}: strip \texttt{BOS}; truncate at first \texttt{EOS}/pad; pad the shorter side; average token-wise equality.
\item \textbf{Normalized Levenshtein similarity (\%)}:
$$100\times \left(1 - \frac{d_L(\hat{s},s)}{\max(1,\max\{|\hat{s}|,|s|\})}\right)$$ where $d_L$ is Levenshtein distance~\cite{levenshtein1966binary}.
\end{itemize}
For $|V|{=}86$, chance level token accuracy is $\approx 1/|V|\approx 1.16\%$.
\begin{tcolorbox}[title=Zero-Shot Decoder Non-Transferability (ZSDN), colback=gray!5!white,colframe=black!75!black]
\textbf{Setting.} Let $\mathcal{F} = \{ f_j \}_{j=1}^n$ be a family of Transformer encoder--decoder models sharing the same public specification $\pi$ (architecture, tokenizer, training data) but initialized using different random seeds, yielding distinct parameter sets $\theta_j = (\theta_j^{\text{enc}}, \theta_j^{\text{dec}})$.

\textbf{Latent Representation.} For an input sequence $M$, the encoder of model $f_j$ produces a latent memory $H^L \in \mathbb{R}^{T \times d_{\text{model}}}$ by stacking $L$ final encoder layer outputs.

\textbf{ZSDN Property.} The system satisfies ZSDN if,
\[
\Pr\big[ f_{j'}^{\text{dec}}\left(H^L;\theta_{j'}^{\text{dec}}\right) =
M \big]
\begin{cases}
\approx \text{chance level}, & \text{if } j'\neq j,\\
\gg \text{chance level}, & \text{if } j' = j.
\end{cases}
\]

\textbf{Decoder-Binding Advantage.} Define
$
\mathsf{Adv}_{\text{bind}} = \text{Acc}_{\text{self}} - \text{Acc}_{\text{cross}},
$
where $\text{Acc}_{\text{self}}$ and $\text{Acc}_{\text{cross}}$ denote token-level accuracy under self- and cross-decoding, respectively. In our results, $\mathsf{Adv}_{\text{bind}} \approx 98\% - 1\% \approx 97\%$.
\end{tcolorbox}
\subsection{Communication and Threat Model}
We assume a setting where the legitimate transmitter $f^j$ publishes $M$'s latent representation $H^L$ over a public channel, akin in spirit to cryptographic encapsulation\cite{goldwasser1984probabilistic}. The legitimate receiver uses the paired decoder $f^{\text{dec}}_j$ to reconstruct the original message $M$ from received $H^L$. The public specification $\pi$ (architecture, tokenizer, training recipe) is known to all parties, while the learned parameters $\theta = (\theta_{\text{enc}}, \theta_{\text{dec}})$ remain private to each model instance. An adversary may:
\begin{itemize}
    \item Observe $(H^L)$. Note that the encoder is operated privately by the transmitting device in this scheme, so the adversary cannot collect $(M, H^L)$ pairs to conduct a (known-plaintext) attack unless messages follow patterns, or query the encoder with chosen inputs (chosen-plaintext).
    \item Attempt zero-shot decoding using a mismatched decoder $f^{j'}_{\text{dec}}$ trained under the same $\pi$ but different seed.
    \item Attempt to learn an adapter or surrogate decoder given limited $(M, H^L)$ pairs.
    \item Attempt to spoof an $\tilde H^L$ encoding an $\tilde M$ as if it was encoded by $f^{\text{enc}}_j$. 
\end{itemize}

Our security goal is \emph{decoder-binding in the zero-shot setting}: without access to $\theta_j^{\text{dec}}$, the adversary's success probability in reconstructing $M$ from $H^L$, as well as spoofing an $H^L$ that is semantically relevant $\tilde M$, should remain near chance. We discuss learnability risks (e.g., adapter alignment) and propose mitigations such as integrity tags, and rekeying in \cref{sec:conc}.

\section{Results}\label{sec:results}

\subsection{Training Dynamics}
All models converged rapidly on the identity mapping task. Final training NLL after $8$ epochs was $0.204$ (M1), $0.170$ (M2), and $0.203$ (M3), decreasing from initial values near $4.05$.\footnote{Per-epoch traces: M1 $4.076\!\to\!0.204$, M2 $4.034\!\to\!0.170$, M3 $4.068\!\to\!0.203$.} This establishes comparable reconstruction capacity across independently initialized replicas.

\subsection{Self- vs. Cross-Decoding Accuracy}
We evaluate all ordered (encoder$\to$decoder) pairs with greedy decoding on held-out batches and report exact sequence match, token accuracy, and normalized Levenshtein similarity in  \cref{tab:decode}. Results show a sharp separation between self-decoding and cross-decoding:
Cross-decoding token accuracy hovers near chance, without exact matches and low Levenshtein similarity. Self-decoding exceeds $91\%$ exact and $98\%$ token accuracy. \texttt{M1\_SAMESEED} and \texttt{M1\_CLONE} reproduce \texttt{M1}’s scores, reflecting identical parameters.

\begin{table}[htbp]
\centering
\caption{Decoding accuracy for representative encoder$\to$decoder pairs.}
\label{tab:decode}
\begin{tabular}{lccc}
\toprule
\textbf{Encoder$\to$Decoder} & \textbf{Exact (\%)} & \textbf{Token (\%)} & \textbf{LevSim (\%)}\\
\midrule
M1$\to$M1           & 91.80 & 98.82 & 99.39 \\
M1$\to$M2           & 0.00  & 1.03  & 4.18  \\
M1$\to$M3           & 0.00  & 0.98  & 3.41  \\
M1$\to$M1\_CLONE    & 91.80 & 98.82 & 99.39 \\
M1$\to$M1\_SAMESEED & 91.80 & 98.82 & 99.39 \\
\midrule
M2$\to$M2           & 88.54 & 98.49 & 99.30 \\
M3$\to$M3           & 86.46 & 97.88 & 98.94 \\
\bottomrule
\end{tabular}
\end{table}

\subsection{Parameter Proximity and Attention Divergence}
We report the $\ell_2$ distance in weight space between model pairs to quantify divergence induced solely by seed differences in \cref{tab:attn-div}. \texttt{M1\_CLONE} is (as expected) identical to M1; \texttt{M1\_SAMESEED} also matched M1 bit-for-bit on this hardware/software configuration, yielding identical decoding behavior. Distinct seeds diverge substantially in parameter space.

\begin{table}[htbp]
{\centering
\caption{Encoder attention divergence (layer-0, head-averaged).}
\label{tab:attn-div}
\begin{tabular}{lccc}
\toprule
\textbf{Pair} & $D_{\ell_2}(\text{A},\text{B})$ & \textbf{KL$(A\Vert B)$} & \textbf{Cosine}\\
\midrule
M1 vs M2            & 324.72 & 0.0961 & 0.8995 \\
M1 vs M3            & 326.71 & 0.0996 & 0.9018 \\
M1 vs M1\_CLONE     & 0.0000 & 0.0000 & 1.0000 \\
M1 vs M1\_SAMESEED  & 0.0000 & 0.0000 & 1.0000 \\
\bottomrule
\end{tabular}}

\end{table}
Mechanism behind cross-decoding failure is quantified by comparing encoder layer-0 head-averaged self-attention maps on identical inputs using KL divergence and cosine similarity. Distinct seeds yield non-zero KL and sub-unity cosine, confirming materially different token-to-token attention distributions despite identical architectures and data. Clone/same-seed pairs are indistinguishable ($\mathrm{KL}{\approx}0$, cosine ${=}1$). Qualitative maps (encoder/decoder self-attention, and decoder cross-attention under true cross-decoding) visually mirror these statistics.
\subsection{Summary of Findings}
Across all evaluations, decoding succeeds \emph{only} when the encoder and decoder parameters are \emph{identical}. Any seed-induced deviation renders the latent $H^L$ effectively undecodable by other models, driving exact match to $0\%$ and token accuracy to chance, illustrating the Anna Karenina principle: success requires all compatibility conditions, while failure results from any misalignment. The attention analyses corroborate that independently trained replicas learn distinct alignment structures, supporting the interpretation of Transformer weights as an implicit, high-dimensional “key” governing decodability.

\begin{figure}[t]
\centering
\includegraphics[width=0.95\linewidth]{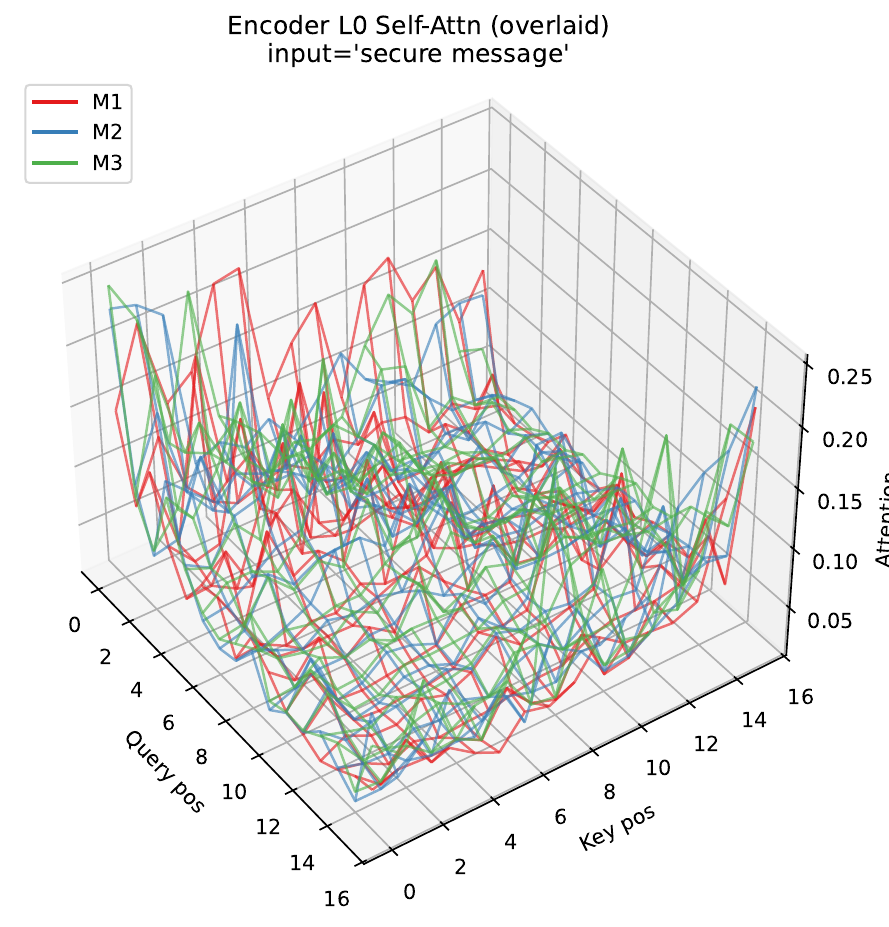}
\caption{Overlaid head-averaged encoder layer-0 self-attention surfaces for the same input (“secure message”) across M1/M2/M3.}
\label{fig:attn_overlay_M1M2M3_encoder}
\end{figure}

The surface shown for model \cref{fig:attn_overlay_M1M2M3_encoder} is the head-average $\bar A_j=\tfrac{1}{h}\sum_{i=1}^{h} A^{(0,i)}_{j}$; the horizontal
axes index query position $q$ and key position $k$, and the vertical axis is the
attention probability $\bar A_j[q,k]$ (each row sums to $1$).
The three colored wireframes correspond to independently trained models
{M1, M2, M3} that share architecture and data but differ in random seeds and
therefore in learned projections $\Theta^{Q,K,V}_j$.

\emph{Why this matters:}  All three encoders exhibit a diagonal bias (local context)
induced by the task and positional encoding, but the locations and magnitudes of
peaks/valleys differ across $j$.
These seed-specific kernels imply different information-mixing operators in the encoder and
therefore different latent geometries $H^L_j=f^{\mathrm{enc}}_j(M;\theta^{\mathrm{enc}}_j)$.
This can be quantified by non-zero distributional distances, e.g.,
$D_{\mathrm{KL}}(\bar A_j\,\|\,\bar A_{j'})>0$ and cosine similarity $<1$ between
flattened maps.  Because each decoder is calibrated to its \emph{own} encoder’s mixing
pattern, swapping encoders/decoders yields a key-mismatch: the latent $H^L_j$ is not
interpretable by $f^{\mathrm{dec}}_{j'}$ for $j'\neq j$, leading to the observed
cross-decoding failure.  Thus, the figure provides direct evidence that encoder
attention acts as a model-specific “key” and illustrates the non-transferability property, which we explore as a candidate for future cryptographic formalization.

\section{Conclusion}\label{sec:conc}

We validated a \emph{model-binding} construction inspired by cryptographic principles informing future cryptographic primitives in which a Transformer autoencoder’s parameters $\theta$ act as the (private) key for a pair of maps
\[
\theta^{\mathrm{enc}}:\ {M}\to \mathcal{C},\qquad 
\theta^{\mathrm{dec}}:\ \mathcal{C}\to {M},
\]
$C$ being the final hidden state $(H^L)$ representing the cipher-text with public algorithmic description $\pi$ (architecture, tokenizer, training recipe), and decryption is performed by the paired decoder. Correctness holds as $\theta^{\mathrm{dec}}(\theta^{\mathrm{enc}}(M))\approx M$ under greedy decoding; \emph{soundness} against model-mismatch is observed as
\[
\theta'^{\mathrm{dec}}(\theta^{\mathrm{enc}}(M)) \ \text{fails for}\ \theta'\neq\theta,
\]
collapsing to chance-level token accuracy, without exact-sequence matches across all tested cross-decoding pairs. Models with identical weights (via cloning or same seed) reproduce self-decoding performance, consistent with a keyed construction. The latent representation from one transformer is effectively in a random basis from the other’s perspective. 

\paragraph*{What fails (and why)}
Despite broadly similar \emph{first-layer} head-averaged encoder attention statistics between independently trained models (small $\mathrm{KL}$, high cosine similarity; cf. \Cref{sec:results}), cross-decoding fails catastrophically. This indicates that the fragile alignment needed for decoding is a property of the \emph{entire} stack—joint bases induced by all $W^{Q,K,V,O,(l)}$, layer norms, and FFNs—rather than any single attention map. In other words, $\theta^{\mathrm{enc}}$ and $\theta'^{\mathrm{dec}}$ operate in incompatible latent coordinate systems when $\theta'\neq\theta$, so the decoder interprets $H^L$ in the "wrong basis" and emits near-random sequences. The organized, near-triangular decoder self-attention we visualize across models is explained by the causal mask and the identity objective; it is necessary for autoregression but not sufficient for successful decoding without the exact parameter alignment.

\paragraph*{Security caveats and hardening}
Unlike number-theoretic ciphers (e.g., RSA or AES), our security rests on parameter non-transferability, not on a reduction to a known hard problem. This non-transferability highlights not just a security primitive, but also a foundation for designing future cryptographic protocols rooted in representation learning. Practical deployments looking to leverage ZSDN today should employ:
\begin{enumerate}
    \item \textbf{Integrity Protection:} Attach a signature or Message Authentication Code (MAC) to $M$ to prevent tampering.
    \item \textbf{Quantization or Noise:} Quantize or inject controlled noise to $H^L$ to reduce leakage and limit oracle attacks.
    \item \textbf{Rekeying Schedule:} Periodically retrain or tune layers to rotate the implicit key, analogous to forward secrecy.
    \item \textbf{Access Control:} Restrict encoder and decoder weights to trusted endpoints; treat $\theta$ as private key material to protect against adapters\cite{pmlr-v97-houlsby19a}, low-rank updates \cite{hu2021loralowrankadaptationlarge}, or prompt/prefix-style conditioning \cite{lester-etal-2021-power}.
    \item \textbf{Rate Limiting:} Limit the number of queries to mitigate chosen-plaintext or model-extraction attacks\cite{Tramer2016ModelStealing,orekondy2019knockoff}.
    \item \textbf{Audit and Logging:} Maintain logs of latent exchanges for anomaly detection and forensic analysis.
\end{enumerate}
Formalizing $\theta^{\mathrm{enc}}$ as a keyed transform and assessing its indistinguishability under $\theta$ drift constitute important future work. Extending ZSDN to larger models and varied tasks will clarify whether the effect can scale into a general security guarantee.

\paragraph*{Takeaway}
Treating learned weights as implicit keys yields a lightweight, accelerator-friendly mechanism for secure intermodel communication without relying on traditional number-theoretic hardness assumptions. This opens the door to secure interoperability protocols between AI agents: correctness for $\theta'=\theta$ and empirical key-mismatch resistance for $\theta'\neq\theta$. This \emph{model-keyed} view aligns with priorities in cryptographic agility and secure AI deployment, enabling provenance tracking, and tamper-resistant inference pipelines. These findings suggest that decoder-binding may serve as a foundation for future cryptographic primitives that jointly leverage representation learning and cryptographic design.

\section*{ACKNOWLEDGMENT}
This material is based upon work supported by the National Science Foundation award CNS-2244515 and the Embry-Riddle Aeronautical University Office of Undergraduate Research. 
Portions of this manuscript were augmented using Microsoft 365 Copilot Researcher and Writing Coach Agents. The final content was reviewed and confirmed by the authors.

\bibliographystyle{jabbrv_IEEEtran}

\bibliography{references}
    
\end{document}